\documentclass[apjl]{emulateapj}
\usepackage{comment}
\usepackage{ifthen}
\usepackage{amsmath}
\usepackage{amsfonts}
\usepackage{amssymb}

\newcommand{\ctbd}[1]{}

\newcommand{\lc}{light curve}
\newcommand{\lcs}{light curves}

\newcommand{\band}[1]{\ensuremath{#1}~band}

\newcommand{\kms}{\ensuremath{\rm km\,s^{-1}}}
\newcommand{\ms}{\ensuremath{\rm m\,s^{-1}}}

\newcommand{\ergscmsq}{\ensuremath{\rm erg\,s^{-1}\,cm^{-2}}}

\newcommand{\vsini}{\ensuremath{v \sin{i}}}
\newcommand{\feh}{\ensuremath{\rm [Fe/H]}}

\newcommand{\vmac}{\ensuremath{v_{\rm mac}}}
\newcommand{\vmic}{\ensuremath{v_{\rm mic}}}
\newcommand{\rsun}{\ensuremath{R_\sun}}
\newcommand{\msun}{\ensuremath{M_\sun}}
\newcommand{\lsun}{\ensuremath{L_\sun}}

\newcommand{\rstar}{\ensuremath{R_\star}}
\newcommand{\mstar}{\ensuremath{M_\star}}
\newcommand{\lstar}{\ensuremath{L_\star}}

\newcommand{\teffstar}{\ensuremath{T_{\rm eff\star}}}
\newcommand{\rhostar}{\ensuremath{\rho_\star}}
\newcommand{\loggstar}{\ensuremath{\log{g_{\star}}}}

\newcommand{\rpl}{\ensuremath{R_{p}}}
\newcommand{\mpl}{\ensuremath{M_{p}}}

\newcommand{\arstar}{\ensuremath{a/\rstar}}

\newcommand{\mjup}{\ensuremath{M_{\rm J}}}

\newcommand{\reffigl}[1]{Figure~\ref{fig:#1}}
\newcommand{\refsecl}[1]{\mbox{Section \ref{sec:#1}}}

\newcommand{\reftabl}[1]{Table~\ref{tab:#1}}

\newcommand{\hatcurhtr}{HTR241-007}                                    % Original HTR name of target
\newcommand{\hatcurfield}{241}                                         % Original HTR field
\newcommand{\hatcurCCra}{\ensuremath{18^{\mathrm h}06^{\mathrm m}09.00{\mathrm s}}}                                  % Right Ascension
\newcommand{\hatcurCCdec}{\ensuremath{+26{\arcdeg}25{\arcmin}36.0{\arcsec}}}                                 % Declination
\newcommand{\hatcurCCtwomass}{2MASS~18060904+2625359}                  % 2MASS identifier
\newcommand{\hatcurCCgsc}{GSC~2099-00908}                              % GSC(1.2) identifier
\newcommand{\hatcurCCtassmv}{11.660}                                   % TASS V-band magnitude
\newcommand{\hatcurCCtwomassJmag}{\ensuremath{10.423\pm0.023}}         % 2MASS ORIG MAG
\newcommand{\hatcurCCtwomassHmag}{\ensuremath{10.128\pm0.027}}         % 2MASS ORIG MAG
\newcommand{\hatcurCCtwomassKmag}{\ensuremath{10.083\pm0.021}}         % 2MASS ORIG MAG
\newcommand{\hatcurCCesoJKmag}{\ensuremath{0.364\pm0.034}}             % 2MASS ESO JK COLOR
\newcommand{\hatcurLCdip}{\ensuremath{5.1}}                            % BLS detected dip (mmag)
\newcommand{\hatcurLCdur}{\ensuremath{0.2161\pm0.0192}}                % transit duration (days)
\newcommand{\hatcurLCdurhr}{\ensuremath{5.187\pm0.462}}                % transit duration (hours)
\newcommand{\hatcurLCq}{\ensuremath{0.0432\pm0.0038}}                  % fractional transit duration (days)
\newcommand{\hatcurLCP}{\ensuremath{5.005037\pm0.000243}}              % period (days)
\newcommand{\hatcurLCPshort}{\ensuremath{5.0050}}                      % period (days)
\newcommand{\hatcurLCT}{\ensuremath{2454350.91482\pm0.00765}}          % epoch (BJD)
\newcommand{\hatcurSMEiteff}{\ensuremath{6065\pm100}}                  % Ini SME, stellar effective temperature
\newcommand{\hatcurSMEizfeh}{\ensuremath{0.15\pm0.08}}                 % Ini SME, stellar metallicity
\newcommand{\hatcurSMEizfehshort}{\ensuremath{0.15}}                   % Ini SME, stellar metallicity
\newcommand{\hatcurSMEilogg}{\ensuremath{4.28\pm0.14}}                 % Ini SME, stellar surface gravity
\newcommand{\hatcurSMEivsin}{\ensuremath{0.5\pm0.6}}                   % Ini SME, stellar rotational velocity
\newcommand{\hatcurSMEivmac}{\ensuremath{4.47}}                        % Ini SME, stellar macroturbulence
\newcommand{\hatcurSMEivmic}{\ensuremath{0.85}}                        % Ini SME, stellar microturbulence
\newcommand{\hatcurSMEiiteff}{\ensuremath{6065\pm100}}                 % Final SME, stellar effective temperature
\newcommand{\hatcurSMEiizfeh}{\ensuremath{0.15\pm0.08}}                % Final SME, stellar metallicity
\newcommand{\hatcurSMEiizfehshort}{\ensuremath{0.15}}                  % Final SME, stellar metallicity
\newcommand{\hatcurSMEiilogg}{\ensuremath{4.28\pm0.14}}                % Final SME, stellar surface gravity
\newcommand{\hatcurSMEiivsin}{\ensuremath{0.5\pm0.6}}                  % Final SME, stellar rotational velocity
\newcommand{\hatcurSMEiivmac}{\ensuremath{4.47}}                       % Final SME, stellar macroturbulence
\newcommand{\hatcurSMEiivmic}{\ensuremath{0.85}}                       % Final SME, stellar microturbulence
\newcommand{\hatcurTRESgamma}{\ensuremath{-2.40\pm0.03}}               % TRES absolute gamma velocity
\newcommand{\hatcurISOm}{\ensuremath{1.20_{-0.06}^{+0.08}}}            % stellar mass
\newcommand{\hatcurISOr}{\ensuremath{1.30_{-0.12}^{+0.25}}}            % stellar radius
\newcommand{\hatcurISOage}{\ensuremath{3.1\pm1.1}}                     % stellar age
\newcommand{\hatcurISOMK}{\ensuremath{2.64\pm0.30}}                    % stellar absolute K magnitude
\newcommand{\hatcurISOJK}{\ensuremath{0.31\pm0.11}}                    % J-K color index from isochrones.
\newcommand{\hatcurISOspec}{G0}                                        % stellar spectral type
\newcommand{\hatcurRVK}{\ensuremath{230.4\pm3.7}}                      % RV semi-amplitude [m/s]
\newcommand{\hatcurPPg}{\ensuremath{6.7_{-2.6}^{+12.1}}}               % planetary surface gravity (m/s^2)
\newcommand{\hatcurPPrho}{\ensuremath{0.12_{-0.06}^{+0.63}}}           % planetary density (cgs)
\newcommand{\hatcurPPm}{\ensuremath{2.13_{-0.07}^{+0.11}}}             % planetary mass (M_jup)
\newcommand{\hatcurPPr}{\ensuremath{2.80\pm0.89}}                      % planetary radius (R_jup)
\newcommand{\hatcurPPteff}{\ensuremath{1364_{-68}^{+108}}}             % planetary temperature (K)
\newcommand{\hatcurPPfluxperidim}{\ensuremath{9}}                      % flux @ periastron (CGS) units.
\newcommand{\hatcurXdist}{\ensuremath{314_{-31}^{+59}}}                % distance (pc)
\newcommand{\hatcur}{HAT-P-31}
\newcommand{\hatcurb}{HAT-P-31b}

\newcommand{\hatcurCCtassvi}{\ensuremath{0.67\pm0.17}}                  % TASS V-I
\newcommand{\hatcurSMEversion}{i}                                       % Final SME version:i or ii?
\newcommand{\hatcurSMEteff}{\ifthenelse{\equal{\hatcurSMEversion}{i}}{\hatcurSMEiteff}{\hatcurSMEiiteff}}
\newcommand{\hatcurSMEzfeh}{\ifthenelse{\equal{\hatcurSMEversion}{i}}{\hatcurSMEizfeh}{\hatcurSMEiizfeh}}
\newcommand{\hatcurSMEzfehshort}{\ifthenelse{\equal{\hatcurSMEversion}{i}}{\hatcurSMEizfehshort}{\hatcurSMEiizfehshort}}
\newcommand{\hatcurSMElogg}{\ifthenelse{\equal{\hatcurSMEversion}{i}}{\hatcurSMEilogg}{\hatcurSMEiilogg}}
\newcommand{\hatcurSMEvsin}{\ifthenelse{\equal{\hatcurSMEversion}{i}}{\hatcurSMEivsin}{\hatcurSMEiivsin}}
\newcommand{\hatcurSMEvmac}{\ifthenelse{\equal{\hatcurSMEversion}{i}}{\hatcurSMEivmac}{\hatcurSMEiivmac}}
\newcommand{\hatcurSMEvmic}{\ifthenelse{\equal{\hatcurSMEversion}{i}}{\hatcurSMEivmic}{\hatcurSMEiivmic}}

\newcommand{\hatcurisoshort}{YY}
\newcommand{\hatcurisofull}{Yonsei-Yale (YY)}
\newcommand{\hatcurisocite}{yi:2001}
\newcommand{\hatcurjhkfilset}{ESO}
\newboolean{emulateapj}
\setboolean{emulateapj}{true}

\newboolean{rvtablelong}
\setboolean{rvtablelong}{true}

\newboolean{astroph}
\setboolean{astroph}{true}
\shortauthors{Kipping et al.}
\shorttitle{\hatcur\lowercase{b}\&\lowercase{c}}
\ifthenelse{\boolean{emulateapj}}{
    \newcommand{\titledag}{$\dagger$}
}{
    \newcommand{\titledag}{\dagger}
}

\begin{document}

%% Titlepage
\title{\hatcur\lowercase{b},\lowercase{c}: A Transiting, Eccentric, Hot Jupiter 
and a Long-Period, Massive Third-Body
	\altaffilmark{\titledag}}

%% Authors
\author{
 D.~M.~Kipping\altaffilmark{1,2},
   J.~Hartman\altaffilmark{1},
   G.~\'A.~Bakos\altaffilmark{1},
   G.~Torres\altaffilmark{1},
   D.~W.~Latham\altaffilmark{1},
   D.~Bayliss\altaffilmark{3},
   L.~L.~Kiss\altaffilmark{3},
   B.~Sato\altaffilmark{4},
   B.~B\'eky\altaffilmark{1},
   G\'eza Kov\'acs\altaffilmark{5},
   S.~N.~Quinn\altaffilmark{1},
   L.~A.~Buchhave\altaffilmark{6},
   J.~Andersen\altaffilmark{6},
   G.~W.~Marcy\altaffilmark{7},
   A.~W.~Howard\altaffilmark{7},
   D.~A.~Fischer\altaffilmark{8},
   J.~A.~Johnson\altaffilmark{9},
   R.~W.~Noyes\altaffilmark{1},
   D.~D.~Sasselov\altaffilmark{1},
   R.~P.~Stefanik\altaffilmark{1},
   J.~L\'az\'ar\altaffilmark{10},
   I.~Papp\altaffilmark{10},
   P.~S\'ari\altaffilmark{10},
   G.~F\H{u}r\'{e}sz\altaffilmark{1}
} 
\altaffiltext{1}{Harvard-Smithsonian Center for Astrophysics,
	Cambridge, MA; email: dkipping@cfa.harvard.edu}

\altaffiltext{2}{Dept. of Physics \& Astronomy, University College London,
	Gower St., London, UK}

\altaffiltext{3}{Research School of Astronomy and Astrophysics, The Australian
National University, Weston Creek, ACT, Australia}

\altaffiltext{4}{Department of Earth and Planetary Sciences, Tokyo Institute of 
	Technology, 2-12-1 Ookayama, Meguro-ku, Tokyo 152-8551}

\altaffiltext{5}{Konkoly Observatory, Budapest, Hungary}

\altaffiltext{6}{Niels Bohr Institute, Copenhagen University, Denmark}

\altaffiltext{7}{Department of Astronomy, University of California,
	Berkeley, CA}

\altaffiltext{8}{Department of Astronomy, Yale University, New Haven, CT}

\altaffiltext{9}{Department of Astrophysics, California Institute of Technology,
	Pasadena, CA, USA}

\altaffiltext{10}{Hungarian Astronomical Association, Budapest, 
	Hungary}

\altaffiltext{$\dagger$}{
	Based in part on observations obtained at the W.~M.~Keck
        Observatory, which is operated by the University of California
        and the California Institute of Technology. Keck time has been
        granted by NASA (N167Hr). Based in part on data collected at
        Subaru Telescope, which is operated by the National
        Astronomical Observatory of Japan. Based in part on
        observations made with the Nordic Optical Telescope, operated
        on the island of La Palma jointly by Denmark, Finland,
        Iceland, Norway, and Sweden, in the Spanish Observatorio del
        Roque de los Muchachos of the Instituto de Astrofisica de
        Canarias.
%%
% TODO: Include other observatories if needed, like ESO/HARPS, LCO, ANU,
% etc. Make sure program ids are here. 
}

%% EOF authors

% #####################################################################
%% abstract
\begin{abstract}
%++++++++++++++++++++++++++++++++++++++++++++++++++++++++++++++++++++++
\begin{comment}

	The abstract should contain the following crucial parameters:

	Stellar parameters:
		\hatcurCCgsc	GSC(1.2) identifier
		\hatcurCCtassmv	(TASS) V magnitude
		\hatcurISOspec	Spectral type of star

	Suggested (optional) stellar parameters:
		\hatcurISOr		Stellar radius
		\hatcurISOm		Stellar mass
		\hatcurSMEteff	Stellar effective temperature.
		\hatcurSMEzfeh	Stellar metallicity
		\hatcurXdist	Distance
		\hatcurISOage	Stellar age

	System params:
		\hatcurLCP		Period
		\hatcurLCT		Transit ephemeris
		\hatcurLCdur	Transit duration

	Planetary params:
		\hatcurPPm		MAss
		\hatcurPPr		Radius
		\hatcurPPrho	Density

	Optional planetary params:
		\hatcurRVK		RV semiamplitude
		\hatcurPPg		Surface gravity
		\hatcurPPteff	Effective temperature

	The abstract should highlight any specialty about this planet (core
	mass, eccentric orbit, new type, outlier in any respect, etc.)
		
\end{comment}
%++++++++++++++++++++++++++++++++++++++++++++++++++++++++++++++++++++++

\setcounter{footnote}{10}

We report the discovery of HAT-P-31b, a transiting exoplanet
orbiting the V=\hatcurCCtassmv\ dwarf star \hatcurCCgsc. HAT-P-31b is the
first HAT planet discovered without any follow-up photometry, demonstrating
the feasibility of a new mode of operation for the HATNet project. The 
$2.17$\,$M_J$, $1.1$\,$R_J$ planet has a period $P_b=5.0054$\,days
and maintains an unusually high eccentricity of $e_b = 0.2450\pm0.0045$,
determined through Keck, FIES and Subaru high precision radial velocities.
Detailed modeling of the radial velocities indicates an additional quadratic 
residual trend in the data detected to very high confidence. We interpret this 
trend as a long-period outer companion, HAT-P-31c, of minimum mass 
$3.4$\,\mjup\ and period $\geq2.8$\,years. Since current RVs span less than
half an orbital period, we are unable to determine the properties of HAT-P-31c
to high confidence. However, dynamical simulations of two possible 
configurations show that orbital stability is to be expected. Further, if 
HAT-P-31c has non-zero eccentricity, our simulations show that
the eccentricity of HAT-P-31b is actively driven by the presence of c, making
HAT-P-31 a potentially intriguing dynamical laboratory.

%, transit epoch 
%$\tau_b = 2454315.877 \pm 0.011$ (BJD$_{\mathrm{UTC}}$), and transit duration 
%$T_{1,4} = 24500\pm3900$\,s.  The host star has a mass
%of \hatcurISOm\,\msun, radius of \hatcurISOr\,\rsun, effective
%temperature \hatcurSMEteff\,K, and metallicity $\feh =+0.15\pm0.08$.  
%Without follow-up photometry (due to the near-integer period
%of HAT-P-31b), we use HATNet data alone to constrain the radius of HAT-P-31b
%to be $R_P\geq0.96\pm0.24$\,\rjup\ whilst the mass is obtained through precise 
%radial velocities to be $2.17\pm0.11$\,\mjup. 

%%
\setcounter{footnote}{0}
\end{abstract}

% #####################################################################
%% keywords

% #####################################################################
%% Introduction
\section{Introduction}
\label{sec:introduction}

Transiting extrasolar planets provide invaluable insight into the nature
of planetary systems. The opportunities for follow-up include spectroscopic
inference of an exoplanet's atmosphere \citep{tinetti:2007}, searches for 
dynamical variations \citep{agol:2005,kipping:2009a,kipping:2009b}, and 
characterizing the orbital elements \citep{winn:2011}. Multi-planet systems in 
particular offer rich dynamical interactions and their frequency is key to 
understanding planet formation.

The Hungarian-made Automated Telescope Network
\citep[HATNet;][]{bakos:2004} survey
%In operation since 2003, it has
%now covered approximately ***11\%*** of the sky, searching for TEPs
%around bright stars ***($8\lesssim I \lesssim 12.5$)***.  
for transiting exoplanets (TEPs)
around bright stars operates six wide-field instruments: four at
the Fred Lawrence Whipple Observatory (FLWO) in Arizona, and two on
the roof of the hangar servicing the Smithsonian Astrophysical
Observatory's Submillimeter Array, in Hawaii.  Since 2006, HATNet has
announced and published 30 TEPs \citep[e.g.][]{johnson:2011}.
In this work, we report our thirty-first discovery, around the
relatively bright star \hatcurCCgsc{}. In addition, a long-period
companion is detected through detailed modeling of the radial velocities,
although no transits of this object have been detected or are necessarily
expected.

In \refsecl{obs}, we summarize the detection of the photometric transit
signal and the subsequent spectroscopic observations of \hatcur{} to
confirm the planet.  In \refsecl{analysis}, we analyze the data to
determine the stellar and planetary parameters. Our findings are
discussed in \refsecl{discussion}.

% #####################################################################
\section{Observations}
\label{sec:obs}

As described in detail in several previous papers 
\citep[e.g.][]{bakos:2010,latham:2009}, HATNet employs the following method to 
discover transiting planets: 1. Identification of candidate transiting planets 
based on HATNet photometric observations. 2. High-resolution, low-S/N 
``reconnaissance'' spectra to efficiently reject many false positives. 3. 
Higher-precision photometric observations during transit to refine transit 
parameters and obtain the light curve derived stellar density. 4. 
High-resolution, high-S/N ``confirmation'' spectroscopy to detect the orbital 
motion of the star due to the planet, characterize the host star, and rule-out 
blend scenarios.

In this work, step three is omitted and this is the biggest difference to
the usual HATNet analysis. The detection, and thus verification, of an exoplanet
can be made using step four alone. Indeed, the majority of exoplanets have been
found in this way. Step 1 clearly allows us to intelligently select the most
favorable targets for this resource-intensive activity though. Step 2 follows
the same logic. Step 3 is predominantly for the purposes of characterizing the 
system, and therefore its omission does not impinge on the planet detection. In
some cases, follow-up photometry is used to confirm marginal HATNet candidate
detections as well, but this is not the case for the discovery presented in this
work. An additional check on step 4 is that the derived ephemeris is consistent
with that determined photometrically.

We did not obtain high precision photometry for this target as the transits
were not observable from our usual site of choice, FLWO, until at least May 
2012. This is because the transiting planet has a near-integer period and the 
time of transit minimum has now phased into the day-light hours in Arizona 
(i.e. unobservable). Rather than wait until this time, we have decided to 
release this confirmed planet detection to the community so that follow-up 
photometry may be conducted at other sites.

The principal consequence of not having any follow-up photometry is that the 
obtainable precision of the transit parameters is reduced. In the case of
HAT-P-31b, this means that the light curve derived density was less precise
than that determined spectroscopically. In practice then, we reverse the usual
logic and instead of applying a prior on the stellar density from the light
curve, we apply a prior on the light curve from the stellar density. One can see 
that the decision on this will vary from case to case depending upon the transit 
depth and target brightness.

Another issue is that in the past HAT analyses have used the HATNet photometry
for measuring the planetary ephemeris, $P$ and $\tau$, and little else.
All other parameters could be more precisely determined from the follow-up
photometry. As a consequence, we used the External Parameter Decorrelation 
\citep[EPD; see][]{bakos:2010} and Trend Filtering Algorithm 
\citep[TFA; see][]{kovacs:2005} techniques to correct the HATNet photometry, 
which are known to attenuate the apparent transit depth by a small amount. This 
was usually accounted for by including an instrumental blending factor, 
$B_{\mathrm{inst}}$, in the HATNet data, which could be determined by comparing 
the ratio of the HATNet apparent depth and the follow-up photometry depth. 
Without follow-up photometry, $B_{\mathrm{inst}}$ would be unconstrained and so 
an estimation of the planetary radius would be impossible. To avoid this, we 
employ the more computationally demanding and sophisticated technique of 
\emph{reconstructive TFA} \citep{kovacs:2005}. Reconstructive TFA does not 
attenuate the transit depth significantly and thus offers a way of avoiding a 
free $B_{\mathrm{inst}}$ term. Our previous experience with the two modes of 
TFA support this. For example, HAT-P-15b's TFA photometry \citep{kovacs:2010} 
was found to require a blending factor of $B_{\mathrm{inst}} = 0.71 \pm 0.07$. 
In contrast, the reconstructive TFA for the same planet causes 
$B_{\mathrm{inst}} = 0.95\pm0.04$. We find no instance in any previous 
implementation of reconstructive TFA where $B_{\mathrm{inst}}$ would depart from 
unity by more than 2-$\sigma$. We will therefore use the reconstructive TFA in 
this work and conservatively double all uncertainties relating to the depth and 
radius of the planet.

In this paper, we will consequently show that HATNet photometry 
alone is sufficient to constrain the system properties and that future work may 
not always require step 3 (i.e. follow-up photometry).

In the following subsections we highlight specific details of this
procedure that are pertinent to the discovery of \hatcurb{}.

% =====================================================================
%% Photometric detection
\subsection{Photometric Detection}
\label{sec:detection}

The transits of \hatcurb{} were photometrically detected with a combined 
confidence of 6.2-$\sigma$ using the HAT-5 telescope in
Arizona and the HAT-8 telescope in Hawaii.  The region around
\hatcurCCgsc{}, a field internally labeled as \hatcurfield, was
observed between 2007 March and 2007 July, whenever
weather conditions permitted.  In total, we gathered 9205 exposures of 5 minutes
at a 5.5 minute cadence.  769 of these images were
rejected by our reduction pipeline because they produced bad
photometry for a significant fraction of stars. A typical image is found to 
contain approximately 48,000 stars down to $I_{C} \sim 14$. For the brightest
stars in the field, the photometric precision per-image was 3\,mmag.

%% ----------------
\begin{figure}[!ht]
\plotone{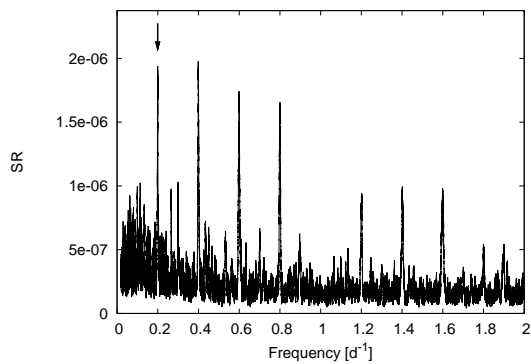} 
\caption{
    Box-fitting Least Squares \citep[BLS;][]{kovacs:2002} periodogram
    of HATNet photometry. The arrow marks the true transit signal
    and the other peaks are interpretted to be aliases. The y-axis
    denotes Signal Residue (SR), see \citet{kovacs:2002} for details
    of the definition.
\label{fig:BLS}}
\end{figure}
%% ----------------

%% ----------------
\begin{figure}[!ht]
\plotone{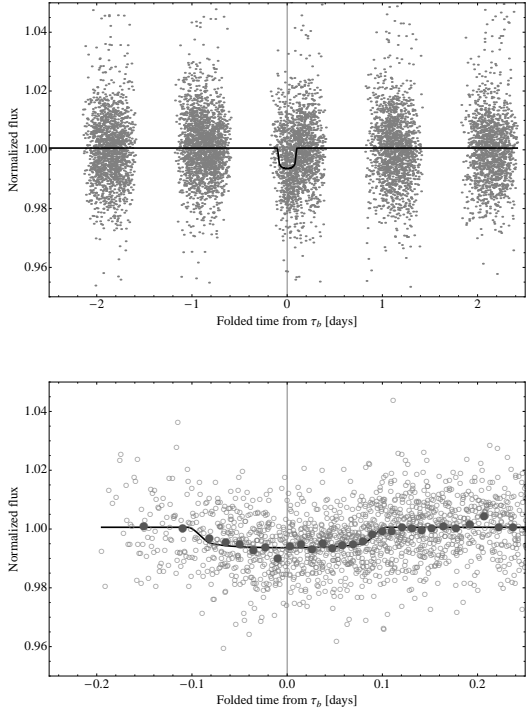} 
\caption{
    Unbinned \lc{} of \hatcur{} including all 8436 instrumental
    \band{I_{C}} 5.5 minute cadence measurements obtained with the
    HAT-5 and HAT-8 telescopes of HATNet (see the text for details),
    and folded with the period $P = 5.005425$\,days resulting
    from the global fit described in \refsecl{analysis}).  The solid
    line shows our transit model fit to the light curve
    (\refsecl{globmod}). The bottom panel shows a zoomed-in view of
    the transit, the filled circles show the light curve binned in
    phase with a bin-size of 50 points.
% Note: the band above can be I, R or Sloan r
%%
\label{fig:hatnet}}
\end{figure}
%% ----------------

Standard photometric procedures were used to calibrate the HATNet frames and
then these calibrated images were subjected to
star detection and astrometry, as described in \cite{pal:2006}. 
Aperture photometry was performed on each image at the stellar
centroids derived from the Two Micron All Sky Survey
\citep[2MASS;][]{skrutskie:2006} catalog and the individual astrometric
solutions.  The resulting \lcs\ were decorrelated (cleaned of trends)
using the External Parameter Decorrelation \citep[EPD; see][]{bakos:2010} 
technique in ``constant'' mode and the Trend
Filtering Algorithm \citep[TFA; see][]{kovacs:2005}. 

The \lcs{} were searched for transits using
the Box-fitting Least-Squares \citep[BLS;][]{kovacs:2002} method.  We detected
a significant signal in the \lc{} of \hatcurCCgsc{} (also known as
\hatcurCCtwomass{}; $\alpha = \hatcurCCra$, $\delta = \hatcurCCdec$;
J2000; V=\hatcurCCtassmv\ \citealp{droege:2006}), with an apparent depth of
$\sim\hatcurLCdip$\,mmag, and a period of $P=\hatcurLCPshort$\,days.  The
BLS periodogram is shown in Figure~\ref{fig:BLS}. The drop 
in brightness had a first-to-last-contact duration, relative to the total 
period, of $q = \hatcurLCq$, corresponding to a total duration of 
$\hatcurLCdurhr$~hr.  Due to the lack of follow-up photometry, the
HATNet photometry was re-processed with more computationally expensive 
reconstructive TFA, as discussed in \S\ref{sec:obs} and shown in 
\reffigl{hatnet}. The EPD and reconstructive TFA corrected photometry is 
provided in Table~\ref{tab:hatnet}.

%% HATNET TABLE
%% --------------------------------------------------------------------
%% 
%\begin{comment}
\begin{deluxetable}{cccc}
\tablewidth{0pc}
\tablecaption{HATNet differential photometry of 
	\hatcur\label{tab:hatnet}}
\tablehead{
	\colhead{BJD} & 
	\colhead{Mag (EPD)\tablenotemark{a}} & 
	\colhead{Mag (TFA)\tablenotemark{b}} &
	\colhead{\ensuremath{\sigma_{\rm Mag}}} \\ 
	\colhead{\hbox{~~~~(2,400,000$+$)~~~~}} & 
	\colhead{} & 
	\colhead{} &
	\colhead{}
}
\startdata
$   54178.0007000 $ & $   11.4164 $ & $   11.4203 $ & $ 0.0075$	\\
$   54178.0045362 $ & $   11.4267 $ & $   11.4273 $ & $ 0.0073$	\\
$   54178.0083829 $ & $   11.4250 $ & $   11.4317 $ & $ 0.0075$	\\
$   54178.0122220 $ & $   11.4177 $ & $   11.4275 $ & $ 0.0069$	\\
$   54178.0160619 $ & $   11.4244 $ & $   11.4220 $ & $ 0.0072$	\\
\vdots & \vdots & \vdots & \vdots \\

[-1.5ex]
\enddata
\tablenotetext{a}{
	These magnitudes have
	been subjected to the EPD procedure.
}
\tablenotetext{b}{
	These magnitudes have
	been subjected to the EPD and TFA procedures.
}
\tablecomments{
    This table is available in a machine-readable form in the online
    journal.  A portion is shown here for guidance regarding its form
    and content.
}
\end{deluxetable}
%% --------------------------------------------------------------------
%\end{comment}

% =====================================================================
\subsection{Reconnaissance Spectroscopy}
\label{sec:recspec}

High-resolution, low-S/N reconnaissance spectra were obtained for
\hatcur{} using the Tillinghast Reflector Echelle Spectrograph
\citep[TRES;][]{furesz:2008} on the 1.5\,m Tillinghast Reflector at
FLWO, and the echelle spectrograph on the Australian National
University (ANU) 2.3\,m telescope at Siding Spring Observatory (SSO)
in Australia. The two TRES spectra of \hatcur{} were obtained, reduced
and analyzed to measure the stellar effective temperature, surface
gravity, projected rotation velocity, and RV via cross-correlation
against a library of synthetic template spectra. The reduction and
analysis procedure has been described by \citet{quinn:2010} and
\citet{buchhave:2010}. A total of 14 spectra of \hatcur{} were obtained
with the ANU 2.3\,m telescope. These data were collected, reduced and
analyzed to measure the RV via cross-correlation against the spectrum
of a RV standard star HD~223311 following the procedure described by
\citet{beky:2011}. The resulting measurements from TRES and the ANU
2.3\,m telescope are given in \reftabl{reconspecobs}.

These observations revealed no detectable RV variation at the
1\,\kms\ precision of the observations. Additionally the spectra are
consistent with a single, slowly-rotating, dwarf star.

%%% TABLE: SUMMARY OF RECONNAISSANCE SPECTROSCOPIC MEASUREMENTS
\begin{deluxetable*}{llccccr}
\tablewidth{0pc}
\tabletypesize{\scriptsize}
\tablecaption{
    Summary of reconnaissance spectroscopy observations of \hatcur{}
    \label{tab:reconspecobs}
}
\tablehead{
    \multicolumn{1}{c}{Instrument}          &
    \multicolumn{1}{c}{Date}                &
    \multicolumn{1}{c}{Number of}           &
    \multicolumn{1}{c}{$\teffstar$}         &
    \multicolumn{1}{c}{$\log(g_*\,[\mathrm{cgs}])$}         &
    \multicolumn{1}{c}{$\vsini$}            &
    \multicolumn{1}{c}{$\gamma_{\rm RV}$\tablenotemark{a}} \\
    &
    &
    \multicolumn{1}{c}{Spectra}             &
    \multicolumn{1}{c}{[K]}                 &
    \multicolumn{1}{c}{}               &
    \multicolumn{1}{c}{[\kms]}              &
    \multicolumn{1}{c}{[\kms]}
}
\startdata
TRES      & 2009 Jul 05 & 1 & $6000$  & $4.0$   & $4$     & $-2.342$ \\
TRES      & 2009 Jul 07 & 1 & $6000$  & $4.0$   & $4$     & $-2.300$ \\
ANU 2.3 m & 2009 Jul 14 & 5 & \nodata & \nodata & \nodata & $-8.07 \pm 0.35$ \\
ANU 2.3 m & 2009 Jul 17 & 5 & \nodata & \nodata & \nodata & $-7.13 \pm 0.40$ \\
ANU 2.3 m & 2009 Jul 18 & 2 & \nodata & \nodata & \nodata & $-7.64 \pm 0.44$ \\
ANU 2.3 m & 2009 Jul 19 & 2 & \nodata & \nodata & \nodata & $-7.54 \pm 0.49$ \\
\enddata 
\tablenotetext{a}{
    The mean heliocentric RV of the target. Systematic differences
    between the velocities from the two instruments are consistent
    with the velocity zero-point uncertainties. For the ANU 2.3\,m
    observations we give the weighted mean of the observations and the
    uncertainty on the mean for each night. Note that the systematic
    difference of $-5.3$\,\kms\ between the ANU 2.3\,m and TRES
    observations is similar to the difference of $-5.1$\,\kms\ found
    between these same two instruments by \cite{beky:2011} for
    HAT-P-27.  
}
\end{deluxetable*}

% =====================================================================
\subsection{High Resolution, High S/N Spectroscopy}
\label{sec:hispec}

\begin{figure*} [ht]
\plotone{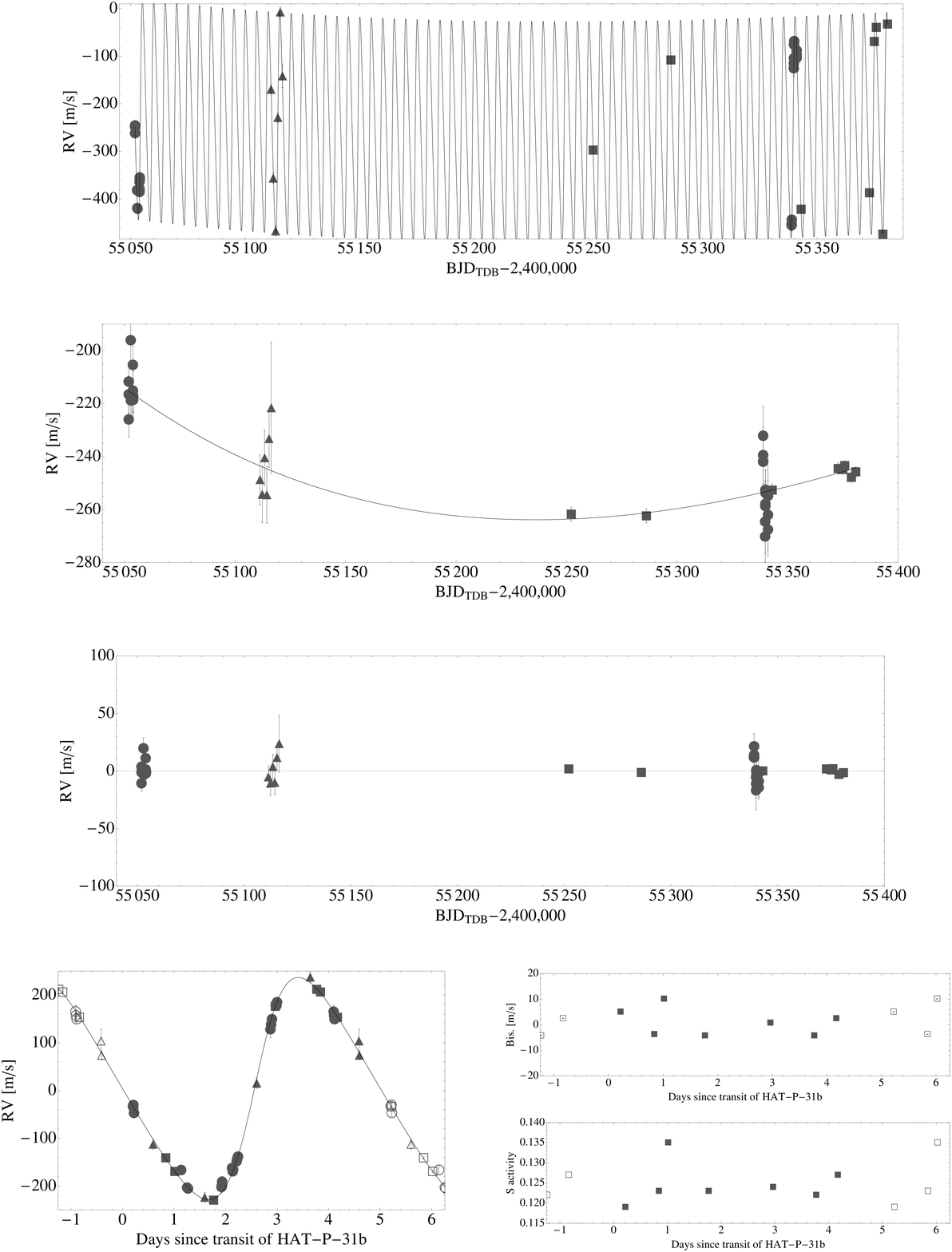}
\caption{
	{\em First Row:} Keck/HIRES (squares), Subaru (circles) and FIES
    (triangles) RV measurements for \hbox{\hatcur{}}, along with our 
    best-fit 2-planet model (see Table~\ref{tab:global}). The 
    center-of-mass velocity has been subtracted.
	{\em Second Row:} Same as top panel except the RV model of the
    inner planet has been subtracted from the data and the model, revealing
    the orbit of the outer planet. The $\chi^2$ of the best-fit 2-planet model
    is 34.8 for 39 data points (rms of 9.60\,m/s) indicating a stellar jitter
    at or below the measurements errors.
	{\em Third Row:} Residuals from our best-fit model.
	{\em Lower left:} RV measurements phased to the orbital periods of the
    inner planet (left). The quadratic trend has been removed.
	{\em Lower right:} Upper panel shows the bisector spans (BS), with the 
    mean value subtracted, phased at the period of the inner planet. Lower
    panel shows relative chromospheric activity index $S$ measured from
    the Keck spectra, phased at the period of the inner planet. Note the 
    different vertical scales of the panels.
    Observations shown twice are represented with open symbols.
\label{fig:rvbis}}
\end{figure*}

We proceeded with the follow-up of this candidate by obtaining
high-resolution, high-S/N spectra to characterize the RV variations,
and to refine the determination of the stellar parameters.  For this
we used the HIRES instrument \citep{vogt:1994} with the iodine-cell
\citep{marcy:1992} on the Keck~I telescope, the High-Dispersion
Spectrograph \citep[HDS;][]{noguchi:2002} with the iodine-cell
\citep{sato:2002} on the Subaru telescope, and the FIbr-fed \'{E}chelle
Spectrograph (FIES) on the 2.5\,m Nordic Optical Telescope
\citep[NOT;][]{djupvik:2010}. \reftabl{hispectab} summarizes the
observations. The table also provides references for the methods used
to reduce the data to relative RVs in the Solar System barycentric
frame. The resulting RV measurements and their uncertainties are
listed in \reftabl{rvs}. The different instrumental uncertainties arise
from different slit widths, exposure times and seeing conditions. The 
period-folded data, along with our best
fit described below in \refsecl{analysis}, are displayed in
\reffigl{rvbis}.

%%% TABLE: SUMMARY OF HIGH-RESOLUTION SPECTROSCOPIC MEASUREMENTS
\begin{deluxetable*}{llcc}
\tablewidth{0pc}
\tabletypesize{\scriptsize}
\tablecaption{
    Summary of high resolution, high S/N spectroscopy observations of \hatcur{}
    \label{tab:hispectab}
}
\tablehead{
    \multicolumn{1}{c}{Instrument}          &
    \multicolumn{1}{c}{Date}               &
    \multicolumn{1}{c}{Number of}           &
    \multicolumn{1}{c}{Reduction}           \\
    &
    \multicolumn{1}{c}{Range}               &
    \multicolumn{1}{c}{RV obs.}             &
    \multicolumn{1}{c}{Reference}           
}
\startdata
HDS       & 2009 Aug 08 -- 2010 May 24 & 25     & 1 \\
HIRES     & 2010 Feb 24 -- 2010 Jul 3  & 9      & 2 \\
FIES      & 2009 Oct 6 -- 2009 Oct 11  & 6      & 3 \\
\enddata 
\tablenotetext{a}{
    The mean heliocentric RV of the target. Systematic differences
    between the velocities from the two instruments are consistent
    with the velocity zero-point uncertainties.
}
\tablerefs{1: \cite{sato:2005}, 2: \cite{butler:1996}, 3: \cite{buchhave:2010}}
\end{deluxetable*}

%% --------------------------------------------------------------------
% FIGURE: RV FIGURE
%% --------------------------------------------------------------------

One false-alarm possibility is that the observed radial velocities are not 
induced by a planetary companion, but are instead caused by distortions in the 
spectral line profiles due to contamination from a nearby unresolved eclipsing 
binary. This hypothesis may be interrogated by examining the spectral line 
profiles for contamination from a nearby unresolved eclipsing binary 
\citep{queloz:2001,torres:2007}. A bisector analysis based on the Keck spectra 
was performed as described in \S5 of \cite{bakos:2007a}. The resulting bisector 
spans, plotted in \reffigl{rvbis}, show no significant variation, and are not 
correlated with the RVs, indicating that this is a real TEP system.

In the same figure, one can also see the S index
\citep{vaughan:1978}, which is a quantitative measure of the chromospheric 
activity of the star derived from the flux in the cores of the \ion{Ca}{2} H
and K lines \citep{isaacson:2010}. Following
\cite{noyes:1984} we find that \hatcur{} has an activity index 
$\log R^{\prime}_{\rm HK} = -5.312$, implying that this is a very 
inactive star.

%
% If using a relative $S$ index and not a calibrated Index
%
\begin{comment}
Note that our relative $S$ index has not been calibrated to the scale
of \citet{vaughan:1978}.  We do not detect any significant variation of
the index correlated with orbital phase; such a correlation might have
indicated that the RV variations could be due to stellar activity.
\end{comment}
%
% If this is true, if not comment on the activity somewhere in the
% paper.
%
\begin{comment}
There is no sign of emission in the cores of the \ion{Ca}{2} H and K
lines in any of our spectra, from which we conclude that the
chromospheric activity level in \hatcur{} is very low.
\end{comment}

%% --------------------------------------------------------------------
%% Note: there are two possible versions for this table: one with
%% BJD RB, RVerr, BS, BSerr, and another one with the former columns
%% plus S and Serr. For the first form it is OK to use the deluxetable 
%% environment. 
%% With the second form we need deluxetable*.
%%
\ifthenelse{\boolean{emulateapj}}{
    \begin{deluxetable*}{lrrrrrr}
}{
    \begin{deluxetable}{lrrrrrr}
}
\tablewidth{0pc}
\tablecaption{
	Relative radial velocities, bisector spans, and activity index
	measurements of \hatcur{}. 
	\label{tab:rvs}
}
\tablehead{
	\colhead{BJD\tablenotemark{a}} & 
	\colhead{RV\tablenotemark{b}} & 
	\colhead{\ensuremath{\sigma_{\rm RV}}\tablenotemark{c}} & 
	\colhead{BS} & 
	\colhead{\ensuremath{\sigma_{\rm BS}}} & 
	\colhead{S\tablenotemark{d}} & 
	\colhead{Instrument}\\
	\colhead{\hbox{[2,454,000$+$]}} & 
	\colhead{[\ms]} & 
	\colhead{[\ms]} &
	\colhead{[\ms]} &
    \colhead{[\ms]} &
	\colhead{} &
	\colhead{}
}
\startdata
%%
%% If the table is too long, then we give only sample lines in the
%% submitted version, and the full table is presented in the electronic
%% version.  As regards the astroph version: in such cases we can
%% decide whether to include all data.
%%
\ifthenelse{\boolean{rvtablelong}}{
    $ 1051.87826 $ & $   -30.72 $ & $     6.99 $ & \nodata      & \nodata      & \nodata      & Subaru \\
$ 1051.88946 $ & $   -27.97 $ & $     6.90 $ & \nodata      & \nodata      & \nodata      & Subaru \\
$ 1051.90067 $ & $   -44.27 $ & $     6.56 $ & \nodata      & \nodata      & \nodata      & Subaru \\
$ 1052.81471 $ & $  -163.76 $ & $     9.25 $ & \nodata      & \nodata      & \nodata      & Subaru \\
$ 1052.93238 $ & $  -200.89 $ & $     6.63 $ & \nodata      & \nodata      & \nodata      & Subaru \\
$ 1052.94706 $ & $  -202.52 $ & $     7.29 $ & \nodata      & \nodata      & \nodata      & Subaru \\
$ 1053.80433 $ & $  -159.83 $ & $     7.09 $ & \nodata      & \nodata      & \nodata      & Subaru \\
$ 1053.81555 $ & $  -166.62 $ & $     7.47 $ & \nodata      & \nodata      & \nodata      & Subaru \\
$ 1053.82676 $ & $  -163.36 $ & $     7.95 $ & \nodata      & \nodata      & \nodata      & Subaru \\
$ 1053.89544 $ & $  -145.48 $ & $     7.08 $ & \nodata      & \nodata      & \nodata      & Subaru \\
$ 1053.90665 $ & $  -140.21 $ & $     6.46 $ & \nodata      & \nodata      & \nodata      & Subaru \\
$ 1053.91786 $ & $  -136.05 $ & $     6.39 $ & \nodata      & \nodata      & \nodata      & Subaru \\
$ 1111.33488 $ & $    71.48 $ & $     9.40 $ & \nodata      & \nodata      & \nodata      & FIES \\
$ 1112.33770 $ & $  -115.00 $ & $    10.70 $ & \nodata      & \nodata      & \nodata      & FIES \\
$ 1113.34016 $ & $  -225.79 $ & $    10.50 $ & \nodata      & \nodata      & \nodata      & FIES \\
$ 1114.34798 $ & $    12.92 $ & $    10.70 $ & \nodata      & \nodata      & \nodata      & FIES \\
$ 1115.38654 $ & $   234.94 $ & $    10.80 $ & \nodata      & \nodata      & \nodata      & FIES \\
$ 1116.33401 $ & $   101.33 $ & $    24.70 $ & \nodata      & \nodata      & \nodata      & FIES \\
$ 1252.10465 $ & \nodata      & \nodata      & $    -2.81 $ & $     1.78 $ & $    0.1280 $ & Keck \\
$ 1252.11375 $ & $   -34.70 $ & $     2.65 $ & $     5.07 $ & $     1.72 $ & $    0.1190 $ & Keck \\
$ 1286.09877 $ & $   153.49 $ & $     2.56 $ & $     2.47 $ & $     1.48 $ & $    0.1270 $ & Keck \\
$ 1338.90673 $ & $  -199.81 $ & $    11.70 $ & \nodata      & \nodata      & \nodata      & Subaru \\
$ 1338.91100 $ & $  -199.43 $ & $    11.83 $ & \nodata      & \nodata      & \nodata      & Subaru \\
$ 1338.91526 $ & $  -196.43 $ & $    10.73 $ & \nodata      & \nodata      & \nodata      & Subaru \\
$ 1338.91953 $ & $  -188.53 $ & $    11.15 $ & \nodata      & \nodata      & \nodata      & Subaru \\
$ 1339.85945 $ & $   129.95 $ & $    16.85 $ & \nodata      & \nodata      & \nodata      & Subaru \\
$ 1339.87200 $ & $   140.14 $ & $    11.90 $ & \nodata      & \nodata      & \nodata      & Subaru \\
$ 1339.88669 $ & $   151.20 $ & $     9.63 $ & \nodata      & \nodata      & \nodata      & Subaru \\
$ 1339.95891 $ & $   180.56 $ & $     7.45 $ & \nodata      & \nodata      & \nodata      & Subaru \\
$ 1339.97360 $ & $   179.56 $ & $     7.50 $ & \nodata      & \nodata      & \nodata      & Subaru \\
$ 1339.98489 $ & $   186.81 $ & $     8.74 $ & \nodata      & \nodata      & \nodata      & Subaru \\
$ 1341.08414 $ & $   167.01 $ & $    13.93 $ & \nodata      & \nodata      & \nodata      & Subaru \\
$ 1341.09189 $ & $   158.61 $ & $    13.28 $ & \nodata      & \nodata      & \nodata      & Subaru \\
$ 1341.10303 $ & $   151.27 $ & $     9.77 $ & \nodata      & \nodata      & \nodata      & Subaru \\
$ 1343.00801 $ & $  -167.85 $ & $     2.30 $ & $    10.12 $ & $     1.17 $ & $    0.1350 $ & Keck \\
$ 1372.86477 $ & $  -139.73 $ & $     2.11 $ & $    -3.69 $ & $     1.30 $ & $    0.1230 $ & Keck \\
$ 1374.99015 $ & $   177.54 $ & $     1.84 $ & $     0.74 $ & $     1.16 $ & $    0.1240 $ & Keck \\
$ 1375.86758 $ & $   207.17 $ & $     1.90 $ & $    -3.49 $ & $     1.11 $ & $    0.1240 $ & Keck \\
$ 1378.79885 $ & $  -228.26 $ & $     1.90 $ & $    -4.23 $ & $     1.29 $ & $    0.1230 $ & Keck \\
$ 1380.79961 $ & $   212.57 $ & $     1.93 $ & $    -4.20 $ & $     1.12 $ & $    0.1220 $ & Keck \\

	[-1.5ex]
}{
    \input{rvtable_short.tex}
	[-1.5ex]
}
\enddata
\tablenotetext{a}{
    Barycentric Julian dates throughout the
    paper are calculated from Coordinated Universal Time (UTC).
}
\tablenotetext{b}{
	The zero-point of these velocities is arbitrary. An overall offset
    $\gamma_{\rm rel}$ fitted to these velocities in \refsecl{globmod}
    has {\em not} been subtracted.
}
\tablenotetext{c}{
	Internal errors excluding the component of astrophysical/instrumental jitter
    considered in \refsecl{globmod}.
}
\tablenotetext{d}{
	Relative chromospheric activity index, not calibrated to the
	scale of \citet{vaughan:1978}.
}
\ifthenelse{\boolean{rvtablelong}}{
	\tablecomments{
		For the iodine-free template exposures there is no RV
		measurement, but the BS and S index can still be determined.
	}
}{
    \tablecomments{
		For the iodine-free template exposures there is no RV
		measurement, but the BS and S index can still be determined.
		This table is presented in its entirety in the
		electronic edition of the Astrophysical Journal.  A portion is
		shown here for guidance regarding its form and content.
	}
} 
\ifthenelse{\boolean{emulateapj}}{
    \end{deluxetable*}
}{
    \end{deluxetable}
}

% #####################################################################
%% Analysis
\section{Analysis}
\label{sec:analysis}
%++++++++++++++++++++++++++++++++++++++++++++++++++++++++++++++++++++++
\begin{comment}
\end{comment}
%++++++++++++++++++++++++++++++++++++++++++++++++++++++++++++++++++++++

The analysis of the \hatcur{} system, including determinations of the
properties of the host star and planet, was carried out in a similar
fashion to previous HATNet discoveries \citep[e.g.][]{bakos:2010}. 
Below, we briefly summarize the procedure and the results for the
\hatcurb{} system.

% =====================================================================
\subsection{Properties of the Parent Star}
\label{sec:stelparam}
%++++++++++++++++++++++++++++++++++++++++++++++++++++++++++++++++++++++
\begin{comment}
	* Stellar atmosphere properties, SME iteration 1.
	  Based on DS or FIES or Keck, etc.
	* Limbdarkening parameters, iteration 1
	* Parallax as luminosity indicator, if available.
	* arstar based luminosity indicator, based on the transit
	* Choice of isochrones
	* SME2, limbdarkening2
	* Distance
	* Stellar properties main table
	* Rotation, activity
\end{comment}
%++++++++++++++++++++++++++++++++++++++++++++++++++++++++++++++++++++++

%%JH_EDIT_START ---- The text between here, and the next occurrence of
%%          JH_EDIT_STOP have been significantly revised.

%% Stellar atmosphere parameters
%%
Stellar atmospheric parameters were measured from our template
Keck/HIRES spectrum using the Spectroscopy Made Easy
\citep[SME;][]{valenti:1996} analysis package, and the atomic line
database of \cite{valenti:2005}.  SME yielded the following values and
uncertainties:
effective temperature $\teffstar=\hatcurSMEiteff$\,K, 
metallicity $\feh=+0.15\pm0.08$\,dex, and 
stellar surface gravity $\loggstar=4.26_{-0.13}^{+0.11}$\,(cgs),
projected rotational velocity $\vsini=\hatcurSMEivsin\,\kms$.

The above atmospheric parameters are then combined with the
\hatcurisofull\ \citep{\hatcurisocite} series of stellar evolution
models to determine other parameters such as the stellar mass, radius
and age. The results are listed in \reftabl{stellar}. We find that the
star has a mass and radius of $\mstar\ = 1.218_{-0.063}^{+0.089}$\,\msun\ and
$\rstar\ = 1.36_{-0.18}^{+0.27}$\,\rsun, and an estimated age of
$3.17_{-1.11}^{+0.70}$\,Gyr.

\begin{figure}[!ht]
\plotone{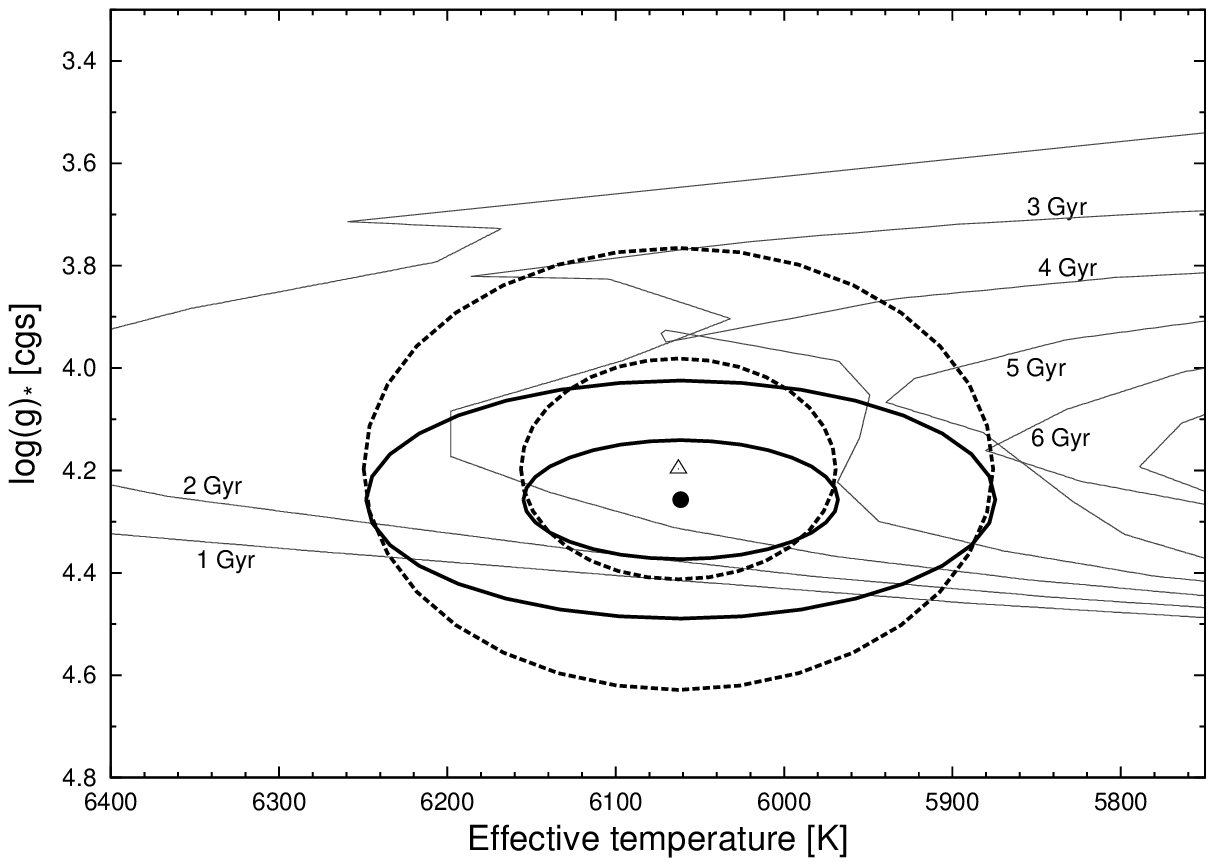}
\caption{
	Model isochrones from \cite{\hatcurisocite} for the measured
        metallicity of \hatcur, \feh = \hatcurSMEiizfehshort, and ages
        from 1 to 14\,Gyr in steps of 1\,Gyr (gray-scale lines, left
        to right).  The adopted values of $\teffstar$ and \loggstar\,
        determined from the SME analysis, are shown together with
        their 1-$\sigma$ and 2-$\sigma$ confidence ellipsoids (filled
        circle, and bold solid lines).  The less-tight 1-$\sigma$ and
        2-$\sigma$ constraints imposed by \arstar\ are also shown
        (open triangle, and bold dotted lines).
\label{fig:iso}}
\end{figure}

For previous HATNet planets \citep[e.g.][]{bakos:2010} we used the
normalized semimajor axis, \arstar\, which is closely related to
\rhostar, the mean stellar density, and is determined from the
analysis of the light curves and RV curves, to obtain an improved
estimate of \loggstar, which is then held fixed in a second SME
iteration. In this case, because we only have the relatively
low-precision HATNet light curve, \arstar\ is poorly constrained, and
we instead opt to use the SME determination of \loggstar\ rather than
\arstar\ as a luminosity indicator. \reffigl{iso} shows the location
of the star in a diagram of \loggstar\ versus \teffstar, together with
the model isochrones. For comparison we also show the relatively poor
constraint on \loggstar\ that is imposed by \arstar.

%% --------------------------------------------------------------------
%% --------------------------------------------------------------------

As an additional check on the stellar evolution modeling, we note that
\hatcur{} has a measured near-infrared color of $J-K =
\hatcurCCesoJKmag$, which we have taken from 2MASS
\citep{skrutskie:2006} using the \citet{carpenter:2001} transformation
to the ESO photometric system. This is within 2-$\sigma$ of the
predicted value from the isochrones of $J-K = \hatcurISOJK$. The
distance listed in \reftabl{stellar} is calculated by comparing the
observed $K$ magnitude (taken from 2MASS and transformed to ESO) to
the absolute $K$ magnitude from the stellar models.

%% --------------------------------------------------------------------
%% Table of stellar parameters. 
%%
\begin{deluxetable}{lrl}
\tablewidth{0pc}
\tabletypesize{\scriptsize}
\tablecaption{
	Stellar parameters for \hatcur{}
	\label{tab:stellar}
}
\tablehead{
	\colhead{~~~~~~~~Parameter~~~~~~~~}	&
	\colhead{Value} &
	\colhead{Source}
}
\startdata
\noalign{\vskip -3pt}
\sidehead{Spectroscopic properties}
\hline
~~~~$\teffstar$ [K]\dotfill         &  \hatcurSMEteff       & SME\tablenotemark{a}\\
~~~~$\feh$\dotfill                  &  \hatcurSMEzfeh       & SME                 \\
~~~~$\vsini$ [\kms]\dotfill         &  \hatcurSMEvsin       & SME                 \\
~~~~$\vmac$ [\kms]\tablenotemark{b}\dotfill          &  \hatcurSMEvmac       & SME                 \\
~~~~$\vmic$ [\kms]\tablenotemark{b}\dotfill          &  \hatcurSMEvmic       & SME                 \\
~~~~$\gamma_{\rm RV}$ [\kms]\dotfill&  \hatcurTRESgamma       & TRES                  \\
\hline
\sidehead{Photometric properties}
\hline
%       Add photometry from other sources, e.g. Tycho-2.
%~~~~$B_T$ [mag]\dotfill            &  10.494 $\pm$ 0.031   & Tycho-2     \\
%~~~~$V_T$ [mag]\dotfill            &  10.038 $\pm$ 0.029   & Tycho-2     \\
~~~~$V$ [mag]\dotfill               &  \hatcurCCtassmv      & TASS                \\
~~~~$V\!-\!I_C$ [mag]\dotfill       &  \hatcurCCtassvi      & TASS                \\
~~~~$J$ [mag]\dotfill               &  \hatcurCCtwomassJmag & 2MASS           \\
~~~~$H$ [mag]\dotfill               &  \hatcurCCtwomassHmag & 2MASS           \\
~~~~$K_s$ [mag]\dotfill             &  \hatcurCCtwomassKmag & 2MASS           \\
%~~~~$E(B\!-\!V)$ [mag]\dotfill     &   0.028 $\pm$ 0.015   & Give the source \tablenotemark{c} \\
%%
\hline
\sidehead{Derived properties}
\hline
~~~~$\mstar$ [$\msun$]\dotfill      &  $1.218_{-0.063}^{+0.089}$      & \hatcurisoshort+SME \tablenotemark{c}\\
~~~~$\rstar$ [$\rsun$]\dotfill      &  $1.36_{-0.18}^{+0.27}$      & \hatcurisoshort+SME         \\
~~~~$\log(g_*\,[\mathrm{cgs}])$\dotfill       &  $4.26_{-0.13}^{+0.11}$       & \hatcurisoshort+SME         \\
~~~~$\lstar$ [$\lsun$]\dotfill      &  $2.23_{-0.58}^{+1.01}$        & \hatcurisoshort+SME         \\
~~~~$M_V$ [mag]\dotfill             &  $3.91_{-0.41}^{+0.34}$         & \hatcurisoshort+SME         \\
~~~~$M_K$ [mag,\hatcurjhkfilset]\dotfill &  \hatcurISOMK    & \hatcurisoshort+SME         \\
~~~~Age [Gyr]\dotfill               &  $3.17_{-1.11}^{+0.70}$        & \hatcurisoshort+SME         \\
~~~~Distance [pc]\dotfill           &  $354_{-51}^{+74}$         & \hatcurisoshort+SME\\
[-1.5ex]
\enddata
\tablenotetext{a}{
	SME = ``Spectroscopy Made Easy'' package for the analysis of
	high-resolution spectra \citep{valenti:1996}.
}
\tablenotetext{b}{Assumed quantity based upon derived spectral type.}
\tablenotetext{c}{
	\hatcurisoshort+SME = Based on the \hatcurisoshort\
    isochrones \citep{\hatcurisocite} and the SME results.
%%
%
% Only if \hatcurlumind = Hip, i.e. we have parallax.
%
%\tablenotetext{d}{The distance given in the table is based on the
%	self-consistent analysis that relies on the Hipparcos parallax and the
%	YY isochrones. It slightly differs from the Hipparcos-based distance.}
%%
}
\end{deluxetable}
%% --------------------------------------------------------------------

%%JH_EDIT_STOP

% =====================================================================
\subsection{Global Modeling of the Data}
\label{sec:globmod}

\subsubsection{Photometry}

In previous HATNet papers, we have used a simplified model for the transit
light curve of the HATNet data. For HAT-P-31, no precise photometry exists
and thus we fit the HATNet data using a more sophisticated quadratic
limb darkening \citet{mandel:2002} algorithm with limb darkening coefficients
interpolated from the tables by \citet{claret:2004}. One caveat is that the 
instrumental blending factor, $B_{\mathrm{inst}}$, is unknown as discussed 
earlier in \S\ref{sec:obs}. We point out that experience with previous HATNet 
planets suggests $B_{\mathrm{inst}}$ is within 2-$\sigma$ of unity for
light curves processed using reconstructive TFA in all cases and thus can be 
accounted for by conservatively doubling the uncertainties on $p$ and $R_P$. 
Further support for a $B_{\mathrm{inst}}$ factor not greatly 
different from unity come from the fact HAT-P-31 is fairly isolated and there 
are no neighbors in 2MASS or a DSS image which contribute significant flux to 
the HATNet aperture. 

Due to the low-precision photometry, the stellar density cannot be determined to 
high precision using the method of \citet{seager:2003} and in fact spectroscopic
estimates were found to be more precise. However, we can reverse this well-known 
trick by implementing a Bayesian prior in our fitting process for the stellar 
density.

We use the spectroscopically determined stellar density from 
\S\ref{sec:stelparam} as a prior in our fits. Since the period of the transiting
planet is well constrained for even low signal-to-noise photometry, reasonably
precise estimates for $P_b$ and $\rho_*$ are possible. With these two 
constrained, $(a_b/R_*)$ is therefore also constrained. The transit light curve
is essentially characterized by four parameters, $\tau_b$, $\delta_b$, $b_b$ and
$(a_b/R_*)$ and thus one of these parameters is constrained by the combination 
of the transit times and the stellar density prior alone.

The photometry which is fitted in the global modeling is corrected for 
instrumental systematics through the EPD and reconstructive TFA correction 
procedures prior to performing the fit (see \S~\ref{sec:detection} and 
\citet{bakos:2010} for details). 

\subsubsection{Radial velocities}

For the radial velocity fits, we found a single planet fit gave a very poor
fit to the observations ($\chi^2 = 204.2$ for 40 RV points) and quickly 
appreciated some kind of second signal was present. Exploring different models,
such as Trojan offsets, polynomial time trends and outer companions,
(see Table~\ref{tab:BICs} for comparison), we found that an eccentric transiting
planet with a quadratic trend in the RVs was the preferred model. The pivot 
point ($t_{\mathrm{pivot}}$) of the polynomial models, including the drift and 
quadratic trends, was selected to be the weighted mean of the radial velocity 
time stamps. 

\begin{figure*}[!ht]
\plotone{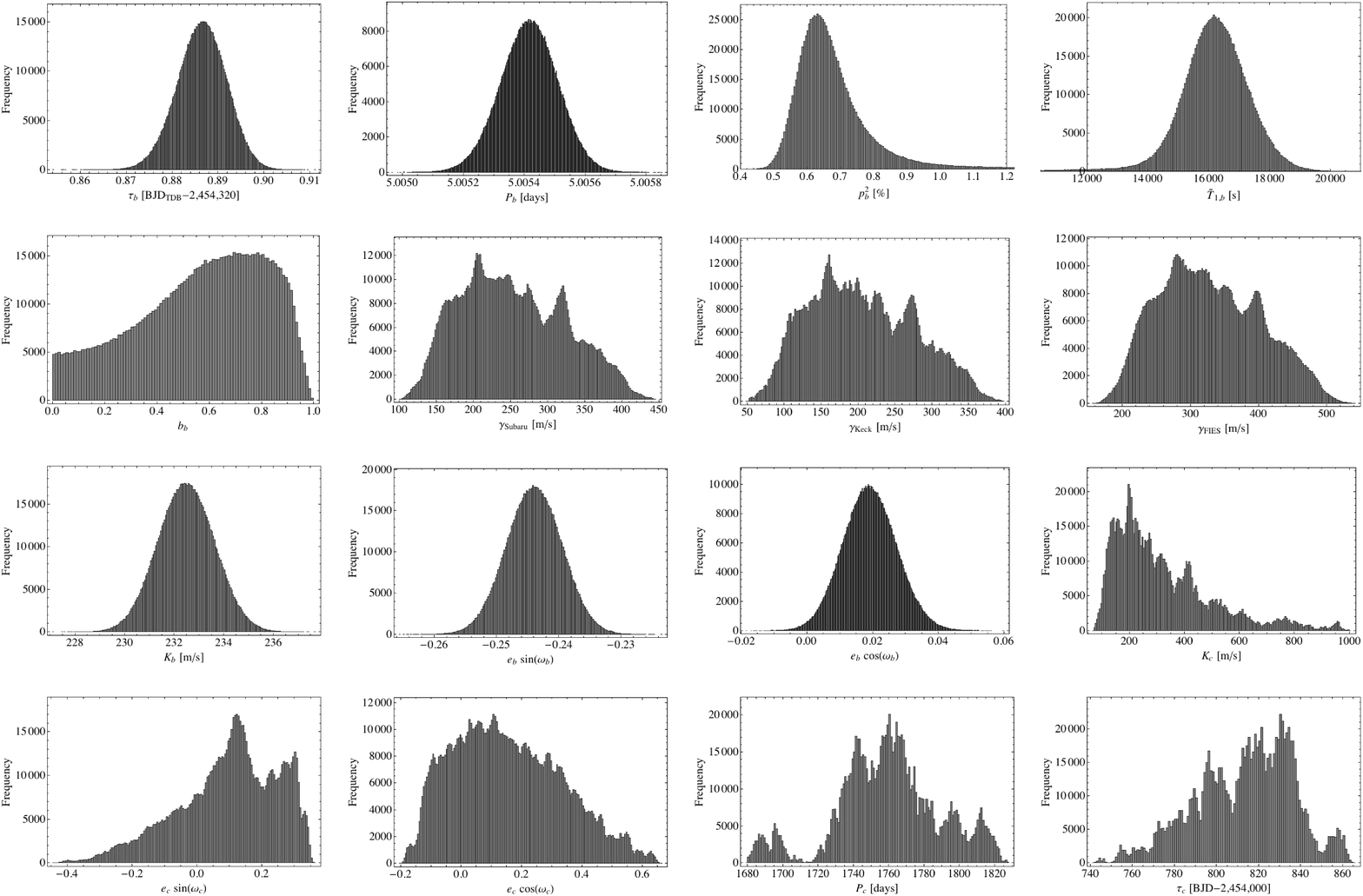}
\caption{
	{\em Top row:} Marginalized posterior distributions for $K_c$ 
        and $P_c$ when we attempted to fit for a second Keplerian
        signal, instead of a quadratic trend. The multi-modal nature
        of these histograms reflect the unconverged nature of the 
        fits.
	{\em Bottom row:} History of the MCMC trials for the same
        parameters and the same fit. Each continuous line represents
        one of the ten independent MCMC chains. The lines clearly
        illustrate the inability of the current data to converge
        upon a solution for HAT-P-31c.
\label{fig:ecceccresults}}
\end{figure*}

The most likely physical explanation for a quadratic trend is a third body in 
the system, described by a Keplerian model. Indeed, the Keplerian model provides 
an improved $\chi^2$ for a circular orbit and then again for an eccentric orbit 
but the extra degrees of freedom penalize our model selection criterion. We also 
found that these models were highly unconstrained and convergence in the 
associated fits was unsatisfactory. An illustration of the lack of convergence
is shown in Figure~\ref{fig:ecceccresults}. Therefore, we will adopt the quadratic model 
in our final reported parameters in Table~\ref{tab:global}.

The quadratic model may be used to infer some physical parameters of the third
planet. To make some meaningful progress, we will assume the outer planet
is on a circular orbit. One may compare the quadratic and Keplerian model
descriptions via:

\begin{align}
\mathrm{RV}_{\mathrm{quad}}^c &= \gamma' + \dot{\gamma} (t - t_{\mathrm{pivot}}) + 0.5 \ddot{\gamma} (t - t_{\mathrm{pivot}})^2 \\
\mathrm{RV}_{\mathrm{Kep}}^c &= \gamma' - K_c \sin\Bigg(\frac{2\pi (t-\tau_c)}{P_c}\Bigg)
\end{align}

By differentiating both expressions and solving for the time when the signals
are minimized, one may write:

\begin{align}
\tau_c + \frac{P_c}{4} &= \frac{ -\dot{\gamma} + \ddot{\gamma}t_{\mathrm{pivot}} }{\ddot{\gamma}}
\label{eqn:obs1}
\end{align}

Differentiating both RV models with respect to time twice
and evaluating at the moment when both signals are minimized, yields:

\begin{align}
\frac{K_c}{P_c^2} &= \frac{\ddot{\gamma}}{4\pi^2}
\label{eqn:obs2}
\end{align}

Equations~\ref{eqn:obs1}\&\ref{eqn:obs2} may therefore be used to determine
some information about HAT-P-31c.

To evaluate the statistical significance of HAT-P-31c, we performed an F-test
between the one-planet and two-planet models. In both cases, HAT-P-31b is
assumed to maintain non-zero orbital eccentricity. Assuming HAT-P-31c is
on a circular orbit, the false-alarm-probability (FAP) from an F-test is
$3.0\times10^{-12}$, or 7.0-$\sigma$. Assuming HAT-P-31c is on an eccentric
orbit requires 2 more degrees of freedom and thus reduces the FAP to
$1.3\times10^{-10}$, or 6.4-$\sigma$. From a statistical perspective then,
the presence of HAT-P-31c is highly secure. We point out this determination
of course assumes purely Gaussian uncertainties and no outlier measurements.

\begin{deluxetable}{lrr}
\tablewidth{0pc}
\tabletypesize{\scriptsize}
\tablecaption{
	Comparison of RV models attempted for HAT-P-31
	\label{tab:BICs}
}
\tablehead{
	\colhead{~~~~~~~~Model~~~~~~~~}	&
	\colhead{$\chi^2$} &
	\colhead{BIC\tablenotemark{a}}
}
\startdata
\noalign{\vskip -3pt}
%%
%\sidehead{Spectroscopic properties} % k = 4!! N = 9 + 25 + 6 for purecirc
Circular Planet\dotfill & 3714.9 & 3729.6 \\
Eccentric Planet\dotfill & 204.2 & 226.3 \\
Eccentric Planet + Drift\dotfill & 159.6 & 185.5 \\
Eccentric Planet + Trojan\dotfill & 191.1 & 216.9 \\
Eccentric Planet + Drift + Trojan\dotfill & 113.2 & 142.7 \\
Eccentric Planet + Drift + Quadratic\dotfill & 33.9 & 63.4 \\
Eccentric Planet + Circular Planet\dotfill & 33.6 & 66.8 \\
Eccentric Planet + Eccentric Planet\dotfill & 33.2 & 73.8 \\
[-1.5ex]
\enddata
\tablenotetext{a}{
	BIC = Bayesian Information Criterion \citep{schwarz:1978,liddle:2007}
}
\end{deluxetable}

\subsubsection{Fitting algorithm}

We utilize a Metropolis-Hastings Markov Chain Monte Carlo (MCMC) algorithm to
globally fit the data, including the stellar density prior (our routine is
described in \citet{kippingbakos:2011}). To ensure the parameter space is fully
explored, we used 5 independent MCMC fits which stop once $1.25\times10^5$ 
trials have been accepted and burn-out the first 20\%. This leaves us with
a total of $5\times10^5$ points for the posterior distributions. At the end of 
the fit, a more aggressive downhill simplex $\chi^2$ minimization is used, for 
which the final solution is used for Figures~\ref{fig:hatnet}\&\ref{fig:rvbis}.

There were 14 free parameters in the global fit: \{$\tau_b$, $p_b^2$, 
$[\tilde{T}_1]_b$, $b_b$, $P_b$, $e_b\sin\omega_b$, $e_b\cos\omega_b$, 
$\gamma_{\mathrm{Keck}}$, $\gamma_{\mathrm{FIES}}$, $\gamma_{\mathrm{Subaru}}$, 
$K_b$, $\dot{\gamma}$, $\ddot{\gamma}$, OOT\}, which we elaborate on here.
$\tau_b$ is the time of transit minimum \citep{thesis:2011}, frequently dubbed 
by the misnomer ``mid-transit time''. $p_b^2$ is the ratio-of-radii squared and 
$P_b$ is the orbital period. $[\tilde{T}_1]_b$ is the ``one-term'' approximate 
equation \citep{investigations:2010} for the transit duration between the 
instant when the center of the planet crosses the stellar limb to exiting under 
the same condition. $b_b$ is the impact parameter, defined as the sky-projected
planet-star separation in units of the stellar radius at the instant of 
inferior conjunction. $e_b$ is the orbital eccentricity and $\omega_b$ is the
associated position of pericenter. $\gamma$ terms relate to the instrumental
offsets for the radial velocities. Similarly, OOT is the out-of-transit flux
for the HATNet photometry. Finally, $K_b$ is the radial velocity semi-amplitude
and $\dot{\gamma}$ (drift) \& $\ddot{\gamma}$ (curl) are the first and second 
time derivatives of $\gamma$.

\begin{figure*}[!ht]
\plotone{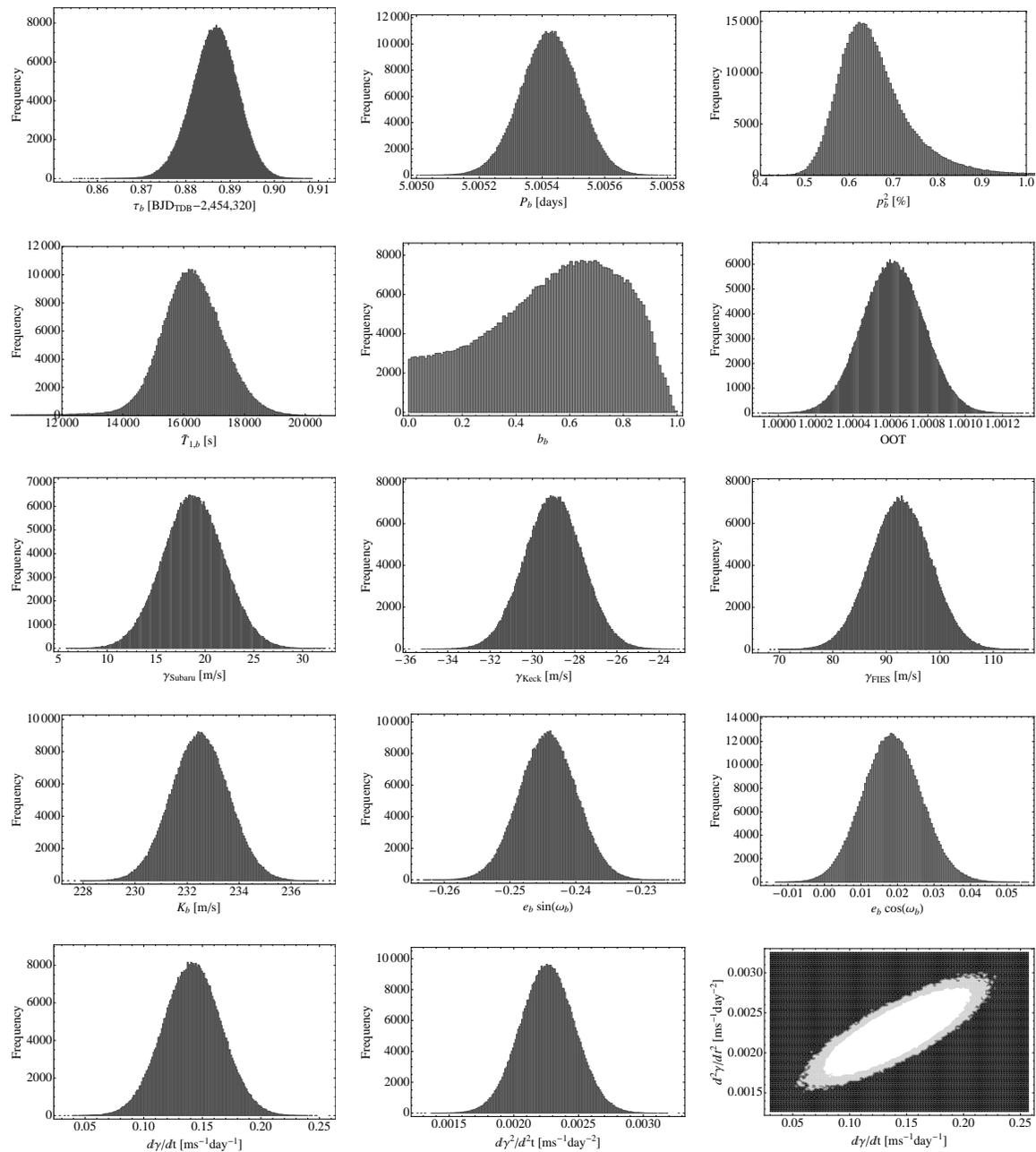}
\caption{
	Posterior distributions of the fitted parameters used in the global
        fits (described in \S\ref{sec:globmod}) from our global fits.
        Histograms computed from $5\times10^5$ MCMC trials. Bottom-right
        panel shows joint-posterior of the radial velocity drift and curl,
        with white denoting the 2-$\sigma$ region of confidence and gray the
        3-$\sigma$. 
\label{fig:histos}}
\end{figure*}

Final quoted results are the median of the marginalized posterior for each
fitted parameter with 34.15\% quantiles either side for the 1-$\sigma$ 
uncertainties (see Table~\ref{tab:global}). The uncertainties on $p$, $p^2$
and $R_P$ have been conservatively doubled for reasons described in 
\S\ref{sec:obs}. Histograms of the posterior
distributions for the fitted parameters are provided in Figure~\ref{fig:histos}.
We find that the stellar jitter of this star is
at or below the measurement uncertainties ($\sim2$\,m/s).

Table~\ref{tab:global} provides estimates for some minimum limits on
various parameters of interest relating to HAT-P-31c. These limits are
determined by the known constraints on the minimum $P_c$. We determined this
value by forcing a circular orbit Keplerian fit for planet c through the data,
stepping through a range of periods from 1\,year to 5\,years in 1\,day steps.
The minimum limit on $P_c$ is defined as when $\Delta\chi^2=1$, relative to
the quadratic trend fit, occurring at 2.8\,years.

%%%%%%%%%%%%%%%%%%%%%%%%%%%%%%%%%%%%%%%%%%%%%%%%%%%%%%%%%%%%%%%%%%%%%%%%%%%%%%%%
%%%%%%%%%%%%%%%%%%%%%%%%%%%%%%%%% SUPER TABLE %%%%%%%%%%%%%%%%%%%%%%%%%%%%%%%%%%
%%%%%%%%%%%%%%%%%%%%%%%%%%%%%%%%%%%%%%%%%%%%%%%%%%%%%%%%%%%%%%%%%%%%%%%%%%%%%%%%

\begin{deluxetable}{lr}
\tablewidth{0pc}
\tabletypesize{\scriptsize}
\tablecaption{
	Global fit results for HAT-P-31
	\label{tab:global}
}
\tablehead{
	\colhead{Parameter\tablenotemark{a}}	&
	\colhead{Value}
}
\startdata
\noalign{\vskip -3pt}
\sidehead{Fitted Parameters}
\hline
$P_b$ [days]\dotfill & $5.005425_{-0.000092}^{+0.000091}$ \\
$\tau_b$ [BJD$_{\mathrm{TDB}}$ - 2,450,000]\dotfill & $4320.8866_{-0.0053}^{+0.0051}$ \\
$[\tilde{T}_1]_b$ [s]\dotfill & $16300_{-1000}^{+1000}$ \\
$p_b^2$ [\%]\dotfill & $0.65_{-0.12}^{+0.18}$ \\
$b_b$\dotfill & $0.57_{-0.31}^{+0.23}$ \\
OOT\dotfill & $1.00061_{-0.00016}^{+0.00016}$ \\
$K_b$ [ms$^{-1}$]\dotfill & $232.5_{-1.1}^{+1.1}$ \\
$e_b\sin\omega_b$\dotfill & $-0.2442_{-0.0043}^{+0.0043}$ \\
$e_b\cos\omega_b$\dotfill & $0.0185_{-0.0079}^{+0.0080}$ \\
$\gamma_{\mathrm{Keck}}$ [ms$^{-1}$]\dotfill & $-29.0_{-1.4}^{+1.4}$ \\
$\gamma_{\mathrm{Subaru}}$ [ms$^{-1}$]\dotfill & $18.8_{-3.1}^{+3.1}$ \\
$\gamma_{\mathrm{FIES}}$ [ms$^{-1}$]\dotfill & $92.7_{-5.6}^{+5.6}$ \\
$\dot{\gamma}$ [ms$^{-1}$day$^{-1}$]\dotfill& $0.141_{-0.025}^{+0.025}$ \\
$\ddot{\gamma}$ [ms$^{-1}$day$^{-2}$]\dotfill& $0.00226_{-0.00021}^{+0.00021}$ \\
\hline
\sidehead{SME Derived Quantities}
\hline
%$T_{\mathrm{eff}}$ [K]\dotfill & $6065 \pm 94$ & $6065 \pm 94$ \\
%$\log(g\,[\mathrm{cgs}])$\dotfill & $4.26_{-0.13}^{+0.11}$ & $4.26_{-0.13}^{+0.11}$ \\
%(Fe/H) [dex]\dotfill & $0.15 \pm 0.08$ & $0.15 \pm 0.08$ \\
$u_1$\tablenotemark{b}\tablenotemark{e}\dotfill & $0.2078^{*}$ \\
$u_2$\tablenotemark{b}\tablenotemark{e}\dotfill & $0.3550^{*}$ \\
%$M_*$ [$M_{\odot}$]\dotfill & $1.218_{-0.063}^{+0.089}$ & $1.218_{-0.063}^{+0.089}$ \\ 
%$R_*$ [$R_{\odot}$]\dotfill & $1.36_{-0.18}^{+0.27}$ & $1.36_{-0.18}^{+0.27}$ \\
$\rho_*$\tablenotemark{c}\tablenotemark{e} [g\,cm$^{-3}$]\dotfill & $0.69_{-0.26}^{+0.34}$ \\
%$L_*$ [$L_{\odot}$]\dotfill & $2.23_{-0.58}^{+1.01}$ & $2.23_{-0.58}^{+1.01}$ \\
%$M_{V}$ [mag]\dotfill & $3.91_{-0.41}^{+0.34}$ & $3.91_{-0.41}^{+0.34}$ \\
%Age [Gyr]\dotfill & $3.17_{-1.11}^{+0.70}$ & $3.17_{-1.11}^{+0.70}$ \\
%Distance [pc]\dotfill & $354_{-51}^{+74}$ & $354_{-51}^{+74}$ \\
\hline
\sidehead{HAT-P-31b Derived Properties}
\hline
$\Psi_b$\dotfill & $0.4737_{-0.0064}^{+0.0065}$ \\
$e_b$\dotfill & $0.2450_{-0.0045}^{+0.0045}$ \\
$\omega_b$ [$^{\circ}$]\dotfill & $274.3_{-1.8}^{+1.8}$ \\
$\log(g_b\,[\mathrm{cgs}])$\dotfill & $3.61_{-0.32}^{+0.15}$ \\
$(a_b/R_*)$\dotfill & $8.9_{-2.3}^{+1.4}$ \\
$i_b$ [$^{\circ}$]\dotfill & $87.1_{-2.7}^{+1.8}$ \\
$[T_{1,4}]_b$ [s]\dotfill & $18500_{-1200}^{+1700}$ \\
$[T_{2,3}]_b$ [s]\dotfill & $14200_{-2400}^{+1300}$ \\
$p_b$\dotfill & $0.080_{-0.015}^{+0.022}$ \\
$M_b$ [$M_J$]\dotfill & $2.171_{-0.077}^{+0.105}$ \\ 
$R_b$ [$R_J$]\dotfill & $1.07_{-0.16}^{+0.24}$ \\
Corr($R_b$,$M_b$)\dotfill & 0.795 \\ 
$\rho_b$ [g\,cm$^{-3}$]\dotfill & $2.18_{-0.93}^{+1.24}$ \\ 
$a_b$ [AU]\dotfill & $0.055_{-0.015}^{+0.015}$ \\
$[T_{\mathrm{eq}}]_b$ [K]\dotfill & $1450_{-110}^{+230}$ \\
$\Theta$\tablenotemark{e} \dotfill & $0.190_{-0.056}^{+0.036}$ \\
$F_{\mathrm{peri}}$ [$10^{\hatcurPPfluxperidim}$\ergscmsq] \dotfill & $1.69_{-0.44}^{+1.39}$ \\
$F_{\mathrm{ap}}$ [$10^{\hatcurPPfluxperidim}$\ergscmsq] \dotfill & $0.62_{-0.16}^{+0.51}$ \\
$\langle F \rangle$\tablenotemark{f} [$10^{\hatcurPPfluxperidim}$\ergscmsq] \dotfill & $0.99_{-0.26}^{+0.82}$ \\
\hline
\sidehead{HAT-P-31c Derived Quantities}
\hline
$\tau_c+0.25 P_c$ [BJD$_{\mathrm{TDB}}$ - 2,450,000]\dotfill & $5254.9_{-6.8}^{+7.4}$ \\
$K_c P_c^{-2}$ [ms$^{-1}$day$^{-2}$]\dotfill & $7.64_{-0.70}^{+0.71}$ \\
$P_c$ [years]\dotfill & $\geq2.8$ \\
$K_c$ [ms$^{-1}$]\dotfill & $\geq60$ \\
$M_c$ [$M_J$]\dotfill & $\geq3.4$ \\
[-1.5ex]
\enddata
\tablenotetext{a}{
Quoted values are medians of MCMC trials with errors given by 1-$\sigma$ 
quantiles. ``b'' subscripts refer to planet HAT-P-31b and ``c'' subscripts 
refer to HAT-P-31c.
}
\tablenotetext{b}{
Fixed parameter
}
\tablenotetext{c}{
Parameter is treated as a prior
}
\tablenotetext{d}{
Parameter determined using SME and YY isochrones
}
\tablenotetext{e}{
	The Safronov number is given by $\Theta = \frac{1}{2}(V_{\rm
	esc}/V_{\rm orb})^2 = (a/\rpl)(\mpl / \mstar )$
	\citep[see][]{hansen:2007}.
}
\tablenotetext{f}{
	Incoming flux per unit surface area, averaged over the orbit.
}
\end{deluxetable}
%%%%%%%%%%%%%%%%%%%%%%%%%%%%%%%%%%%%%%%%%%%%%%%%%%%%%%%%%%%%%%%%%%%%%%%%%%%%%%%%
%%%%%%%%%%%%%%%%%%%%%%%%%%%%%%%%%%%%%%%%%%%%%%%%%%%%%%%%%%%%%%%%%%%%%%%%%%%%%%%%
%%%%%%%%%%%%%%%%%%%%%%%%%%%%%%%%%%%%%%%%%%%%%%%%%%%%%%%%%%%%%%%%%%%%%%%%%%%%%%%%

%% --------------------------------------------------------------------

%% --------------------------------------------------------------------

%% --------------------------------------------------------------------
%% --------------------------------------------------------------------

% #####################################################################
%% Discussion
\section{Discussion}
\label{sec:discussion}

\subsection{Physical Properties of HAT-P-31b\&c}

HAT-P-31b is a $2.17$\,$M_J$ hot-Jupiter transiting the host star once every 
$5.005$\,days. Due to the lack of follow-up photometry obtained for this object
(as a consequence of the nearly integer orbital period), we have only HATNet
photometry, which is of lower signal-to-noise than dedicated follow-up.
This fact, combined with our choice to double all uncertainties on depth
related transit terms, leads to a large uncertainty on the planetary radius of
$R_b = 1.07_{-0.32}^{+0.48}$\,$R_J$, consistent with many other hot-Jupiter objects 
(see http://exoplanet.eu).

High-precision radial velocities also indicate the presence of an outer
body, HAT-P-31c, found through an induced quadratic trend in the RV residuals.
Keplerian fits are unable to convincingly distinguish between a circular or 
eccentric orbit for this object. HAT-P-31c has a minimum mass of 
$M_c\geq3.4$\,$M_J$ and eccentric orbit solutions significantly 
increase this figure. It is unclear whether HAT-P-31c is a brown 
dwarf or a ``planet'', and future work will be needed to determine this.

\subsection{Orbital Stability}
\label{sub:dynamics}

\subsubsection{Circular fit for HAT-P-31c}

Here we discuss our procedure to test the dynamical stability of two 
possible orbital configurations. It should be noted that the eccentricity of 
HAT-P-31b is highly secure but the eccentricity of HAT-P-31c remains unclear.
For this reason, we repeat our simulations assuming both a circular and 
eccentric orbit for HAT-P-31c, beginning with the former. We utilize the 
Systemic Console \citep{meschiari:2009} for this purpose assuming a coplanar 
configuration. Employing the Gragg-Bulirsch-Stoer integrator, orbital evolution 
was computed for 250,000\,years for the HAT-P-31 system (see 
Figure~\ref{fig:evolution}). 

We first consider the circular case. The orbital period and mass of HAT-P-31c
are non-convergent parameters and so we can only provide an orbital solution
which gives a good fit to the data, but is not necessarily unique.
To this end, we proceeded to input HAT-P-31c with $P_c = 4.86$\,years and 
$M_c = 13.0$\,$M_J$, corresponding to the solution presented in 
Table~\ref{tab:BICs}. This test revealed minor evolution over the course of our 
simulations, indicating a stable and essentially static configuration.

\subsubsection{Eccentric fit for HAT-P-31c}

\begin{figure*}[!ht]
\plotone{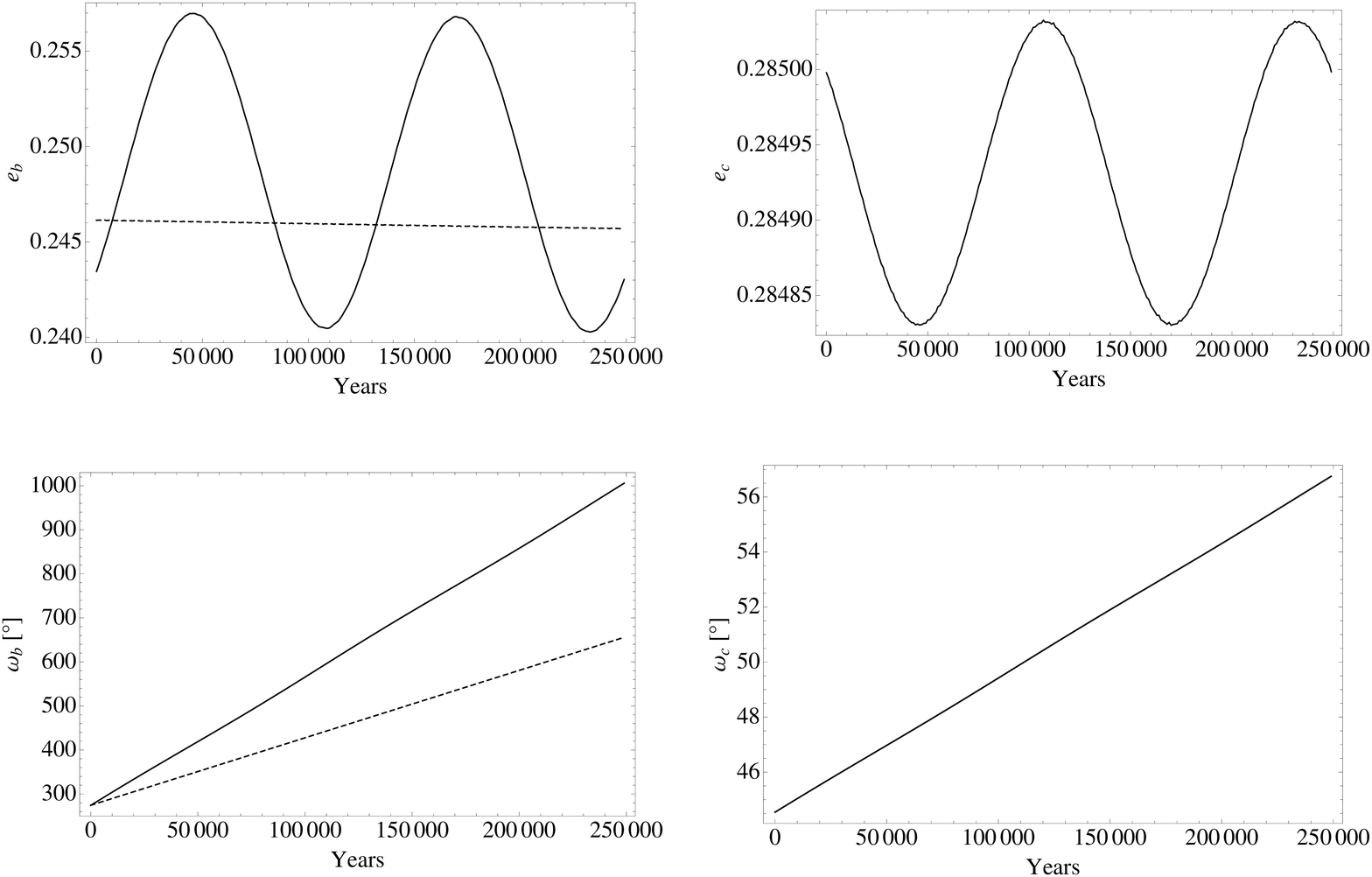}
\caption{
	Two possible realizations for the orbital evolution of planets HAT-P-31b 
	and c. The solid lines show
        the evolution starting from an eccentric orbit solution for HAT-P-31c.
        The dashed lines show the evolution starting from a circular orbit
        solution for HAT-P-31c.
\label{fig:evolution}}
\end{figure*}

To test the eccentric fit, we again used the lowest $\chi^2$ solution
presented earlier in Table~\ref{tab:BICs},
corresponding to $P_c = 4.82$\,years and $e_c = 0.285$. We found
that the system was also stable over 250,000\,years (see 
Figure~\ref{fig:evolution}). However, the simulations do show the eccentricity 
of planet b varying sinusoidally over a timescale of $\sim$125,000\,years with 
an amplitude of $\sim0.01$. The eccentricity of planet c also varies in 
anti-phase but with a much smaller amplitude. The eccentric evolution shows 
faster apsidal precession for planet b, but this is unlikely to be observable 
through changes in the transit duration. We estimate the duration will change by 
0.2\,s over 10\,years (corresponding to $\Delta \omega_b = 0.027^{\circ}$) 
using the expressions of \citet{investigations:2010}.

The orbital period and semi-major axes of both bodies were stable over the
250,000\,years of integration considered here.

\subsubsection{Habitable-zone bodies}

We tried adding a habitable-zone Earth-mass planet on a circular orbit
into the system and testing stability. One may argue that the
probable history of this system involved the inwards migration of
HAT-P-31b and that this migration through the inner protoplanetary
disk would essentially eliminate the possibility of an Earth-like
planet forming in the habitable-zone. However, \citet{fogg:2007}
have shown that this not necessarily true. In their simulations, it is
found that $>60$\% of the solid disk survives, including planetesimals 
and protoplanets, by being scattered by the giant planet into external 
orbits where dynamical friction is strong and terrestrial planet 
formation is able to resume. In one simulation, a planet of 
$2\,M_{\oplus}$ formed in the habitable-zone after a hot-Jupiter 
passed through and its orbit stabilized at 0.1\,AU.

For a planet to receive the same
insolation as the Earth, we estimate $P = 604$\,days. For our circular
orbit solution of HAT-P-31c, the habitable-zone Earth-mass planet is
stable for over 100,000\,years. For our eccentric orbit solution, the Earth-like
planet is summarily ejected in less than 1000\,years.

%% --------------------------------------------------------------------
%% --------------------------------------------------------------------

\subsection{Circularization Timescale}

Due to the poor constraints on the planetary radius, there is a great deal
of uncertainty in the circularization timescale ($\tau_{\mathrm{circ}}$) for 
HAT-P-31b. Nevertheless, using the equations of \citet{adams:2006}, we used the 
MCMC results to compute the posterior distribution of $\tau_{\mathrm{circ}}$.
We find that the age of HAT-P-31 is equal to $24_{-15}^{+122}$ circularization
timescales, assuming $Q_P = 10^5$. This indicates that 
we currently have insufficient data to assess whether the observed eccentricity 
is anomalous or not. Improved radius constraints will certainly aid in this 
calculation and may lend or detract credence to the hypothesis of eccentricity 
pumping of the inner planet by HAT-P-31c.

%\subsection{Tidal Heating of HAT-P-31b}
%
%The eccentric orbit of HAT-P-31b, combined with its short period, makes the 
%planet a candidate for significant tidal heating. The incoming stellar radiation
%hitting the planetary surface is $L_* R_P^2 (1-e^2)^{1/4}/(4a^2)$ and it is
%this value to which any tidal heating should be compared.
%% New text, in reaction to Heller's comments
%The range of possible tidal heating values for HAT-P-31b depend
%upon the obliquity of the planetary orbit, which is currently
%unknown. For the full range of obliquities, 
%%Using the expressions of \citet{peale:1978}, we estimate 
%%$\dot{E}_{\mathrm{tidal}}/\dot{E}_{\mathrm{stellar}} = (2.0_{-1.1}^{+6.2})\times10^{-9}$,
%%indicating that tidal heating is unlikely to be a significant source of energy
%%for HAT-P-31b, despite the large eccentricity.

% #####################################################################
%% Acknowledgements
\acknowledgements 

HATNet operations have been funded by NASA grants NNG04GN74G,
NNX08AF23G and SAO IR\&D grants.  
% Individuals' grants
D.M.K. has been supported by Smithsonian 
Institution Restricted Endowment Funds. Work of G.\'A.B.~and J.~Johnson were
supported by the Postdoctoral Fellowship of the NSF Astronomy and
Astrophysics Program (AST-0702843 and AST-0702821, respectively).  GT
acknowledges partial support from NASA grant NNX09AF59G.  We
acknowledge partial support also from the Kepler Mission under NASA
Cooperative Agreement NCC2-1390 (D.W.L., PI).  G.K.~thanks the
Hungarian Scientific Research Foundation (OTKA) for support through
grant K-60750.  
L.L.K. is supported by the ``Lendulet'' Young Researchers Program
of the Hungarian Academy of Sciences and the Hungarian
OTKA grants K76816, K83790 and MB08C 81013.
Tam\'as Szalai (Univ. of Szeged) is acknowledged for his
assistance during the ANU 2.3 m observations.
This research has made use of Keck telescope time
granted through NASA (N167Hr).Based in part on data collected at
Subaru Telescope, which is operated by the National Astronomical
Observatory of Japan. Based in part on observations made with the
Nordic Optical Telescope, operated on the island of La Palma jointly
by Denmark, Finland, Iceland, Norway, and Sweden, in the Spanish
Observatorio del Roque de los Muchachos of the Instituto de
Astrofisica de Canarias. Thanks to G. Laughlin for useful advise on the 
Systemic Console.
Thanks to Dan Fabrycky and Ren\'e Heller for useful comments. Special
thanks to the anonymous referee for their helpful suggestions.

%% EOF Acknowledgements

% #####################################################################
%% Bibliography
%%%% %T\n %%%% %u\n \\bibitem[%\3m%(y)]{%\1h:%\Y} %\8l~%\Y,%\j,%\V,%\p\n

%% \%% %3.3a\n\%% %t\n\%% %R\n\%%%u\n\\bibitem[%\4m%(y)]{%\1H%\Y} 
%\8l~%\Y,%\j,%\V,%\p\n

\end{document}